\documentclass[twocolumn,aps,prl,longbibliography,sort&compress,showpacs,amsfonts,amsmath,amssymb,superscriptaddress,footinbib,floatfix]{revtex4-1}
\usepackage[T1]{fontenc}
\usepackage[latin9]{inputenc}
\usepackage{amsmath}
\usepackage{amssymb}
\usepackage{graphicx}

\makeatletter
\usepackage{subfigure}
\usepackage{siunitx}

\usepackage{ulem}
\usepackage[dvipsnames]{xcolor}

\usepackage{filemod}

\def\kbar{{\mathchar'26\mkern-9mu k}}
\newcommand{\dd}{\text{d}}

\newcommand{\old}[1]{\textcolor{Red}{\sout{#1}}}

\renewcommand{\old}[1]{}

\makeatother

\begin{document}

\title{Ratchet Effect in the Quantum Kicked Rotor and its Destruction by
Dynamical Localization}

\author{Cl\'ement Hainaut}

\affiliation{Universit\'e de Lille, CNRS, UMR 8523 \textendash{} PhLAM \textendash{}
Laboratoire de Physique des Lasers Atomes et Mol\'ecules, F-59000 Lille,
France}

\author{Adam Ran\c con}

\affiliation{Universit\'e de Lille, CNRS, UMR 8523 \textendash{} PhLAM \textendash{}
Laboratoire de Physique des Lasers Atomes et Mol\'ecules, F-59000 Lille,
France}

\author{Jean-Fran\c cois Cl\'ement}

\affiliation{Universit\'e de Lille, CNRS, UMR 8523 \textendash{} PhLAM \textendash{}
Laboratoire de Physique des Lasers Atomes et Mol\'ecules, F-59000 Lille,
France}

\author{Jean Claude Garreau}

\affiliation{Universit\'e de Lille, CNRS, UMR 8523 \textendash{} PhLAM \textendash{}
Laboratoire de Physique des Lasers Atomes et Mol\'ecules, F-59000 Lille,
France}

\author{Pascal Szriftgiser}

\affiliation{Universit\'e de Lille, CNRS, UMR 8523 \textendash{} PhLAM \textendash{}
Laboratoire de Physique des Lasers Atomes et Mol\'ecules, F-59000 Lille,
France}

\author{Radu Chicireanu}

\affiliation{Universit\'e de Lille, CNRS, UMR 8523 \textendash{} PhLAM \textendash{}
Laboratoire de Physique des Lasers Atomes et Mol\'ecules, F-59000 Lille,
France}

\author{Dominique Delande}

\affiliation{Laboratoire Kastler Brossel, UPMC-Sorbonne Universit\'es, CNRS, ENS-PSL
Research University, Coll\`ege de France, 4 Place Jussieu, 75005 Paris,
France}
\begin{abstract}
We study experimentally a  quantum kicked
rotor with broken parity symmetry, supporting a ratchet effect
due to the presence of a classical accelerator mode.
We show that the short-time dynamics is very well described by the
classical dynamics, characterized by a strongly asymmetric momentum
distribution with directed motion
on one side, and an anomalous diffusion on the other. At longer times,
quantum effects lead to dynamical localization, causing an asymptotic
resymmetrization of the wave function. 
\end{abstract}


\maketitle
Quantum transport phenomena play a central role in many areas in physics,
such as coherently controlled photocurrent in semiconductors \cite{Shapiro:QuantumControlMolecular:12}
or atoms in optical lattices \cite{Morsch:DynamicsBECOptLatt:RMP06}.
Non linearity can lead to interesting effects even for really simple
models, with numerous applications, such as dynamics of Josephson
junctions \cite{Abrikosov:FundamentalsTheoryMetals:72} and electronic
transport through superlattices \cite{Bass:KineticElectrodynamicPhSemiconductors:97},
to cite a few. Symmetries in the system are of fundamental importance,
as their presence or absence can enhance or destroy (quantum) interferences,
and can therefore strongly affect transport properties. In particular,
in  systems with broken parity symmetry, one can observe
directed transport \cite{Weiss:RatchetEffectSQUIDs:EPL00}. The
breaking of time reversal symmetry can also lead to directed transport
\cite{Flach:DirectedCurrentBrokenTimeSpaceSym:PRL00}, an example
being the ratchet effect, that is, the existence
of directed transport without a bias field in periodic systems.

The study of the ratchet effect was  stimulated
by the research on Brownian molecular motors e.g. in biological
systems, and was investigated in different simple noisy models \cite{Reimann:BrownianMotors:PREP02}.
In its original formulation, this effect is of stochastic nature,
due to an external (for instance thermal) noise. A possible variant
are the so-called Hamiltonian chaotic ratchets \cite{Dittrich:ClassicalAndQuantumTransportHRachet:ANPH18,Schanz:ClassicalAndQuantumHamiltonianRatchets:PRL01,Gong:DirectedAnomalousDiffusionBiasedRatchet:PRE04},
where the extrinsic noise is replaced by deterministic chaos, possibly
in presence of dissipation \cite{Jung:RegularAndChaoticTransport:PRL96,Mateos:ChaoticTransportCurrentReversal:PRL00,Porto:MolecularMotorNeverSteps:PRL00,Flach:DirectedCurrentBrokenTimeSpaceSym:PRL00,Gommers:DissipationSymmetryBreakOptLatt:PRL05,Renzoni:QuasiperiodicRatchets:PRL06}.
These ratchets require mixed phase spaces displaying regular
regions embedded in a chaotic sea \cite{Schanz:ClassicalAndQuantumHamiltonianRatchets:PRL01}.
However, it can be shown that in a Hamiltonian chaotic ratchet
the current averaged on the whole phase-space is always zero for unbiased
potentials. The accelerated islands in phase-space contribute to
 directed transport but this effect is globally compensated by
the motion of the remaining part of the phase space, the chaotic
sea \cite{Dittrich:ClassicalAndQuantumTransportHRachet:ANPH18,Schanz:ClassicalAndQuantumHamiltonianRatchets:PRL01}.

The kicked rotor is a paradigm of classical chaotic dynamics,
displaying, according to parameters, a regular, mixed or ergodic phase
space. It is also known to display accelerator modes, with a distinct
ballistic behavior. In its quantum version the Quantum Kicked Rotor
(QKR) is a benchmark model for quantum simulations, intensively in
the context of quantum chaos \cite{Izrailev:LocDyn:PREP90}. Moreover,
it has been shown to display the striking phenomenon of \emph{dynamical
localization} \cite{Casati:LocDynFirst:LNP79,Moore:LDynFirst:PRL94}:
at long times, the diffusive classical dynamics is inhibited by quantum
effects, which have been shown mathematically to be equivalent to the
Anderson localization in momentum space \cite{Fishman:LocDynAnders:PRL82}.
The first realization of the QKR using cold atoms \cite{Moore:AtomOpticsRealizationQKR:PRL95}
has triggered numerous experiments in the field of quantum chaos \cite{DArcy:AccModes:PRL99,Monteiro:DirectedMotionDKR:PRL07,Kenfack:ControllingRatchetEffectColdAt:PRL08,White:ExperimentalQuantumRatchet:PRA13}.
In particular,  adding a quasi-periodic modulation of the kick amplitude
allows one to maps the system onto multidimensional Anderson models
\cite{Shepelyansky:Bicolor:PD87,Casati:IncommFreqsQKR:PRL89}. This
made it possible to study 2D Anderson localization \cite{Manai:Anderson2DKR:PRL15},
and to fully characterize the 3D metal-insulator Anderson transition
\cite{Chabe:Anderson:PRL08,Lemarie:CriticalStateAndersonTransition:PRL10,Lopez:ExperimentalTestOfUniversality:PRL12,Lopez:PhaseDiagramAndersonQKR:NJP13,Lemarie:AndersonLong:PRA09}.

In this work we use the cold-atom realization of the Kicked
Rotor to, i)  demonstrate that directed motion can be generated if
the system's parity invariance is broken, ii)  characterize the classical anomalous diffusion of the chaotic sea, and iii) show that the subtle
quantum interferences at the origin of dynamical localization have
the striking effect of counteracting this directed classical transport.
At long times, the transport asymmetry associated with the ratchet dynamics
disappears and the system undergoes dynamical localization, associated
with a symmetrical wave function with a characteristic universal shape.

\begin{figure*}[t!]
\begin{centering}
\includegraphics[width=0.45\linewidth]{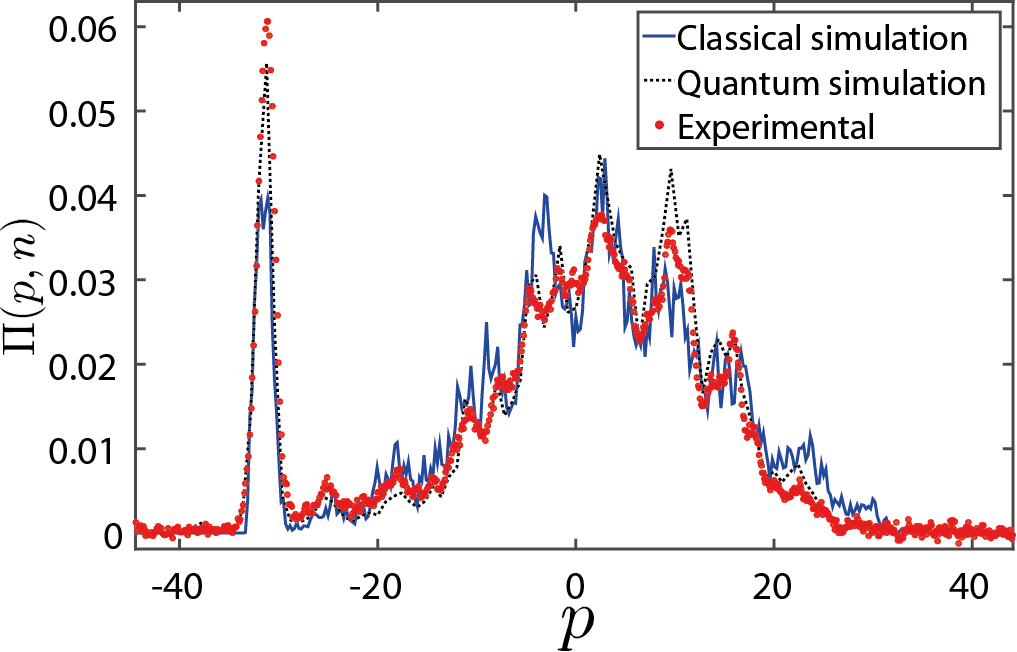} \includegraphics[width=0.43\linewidth]{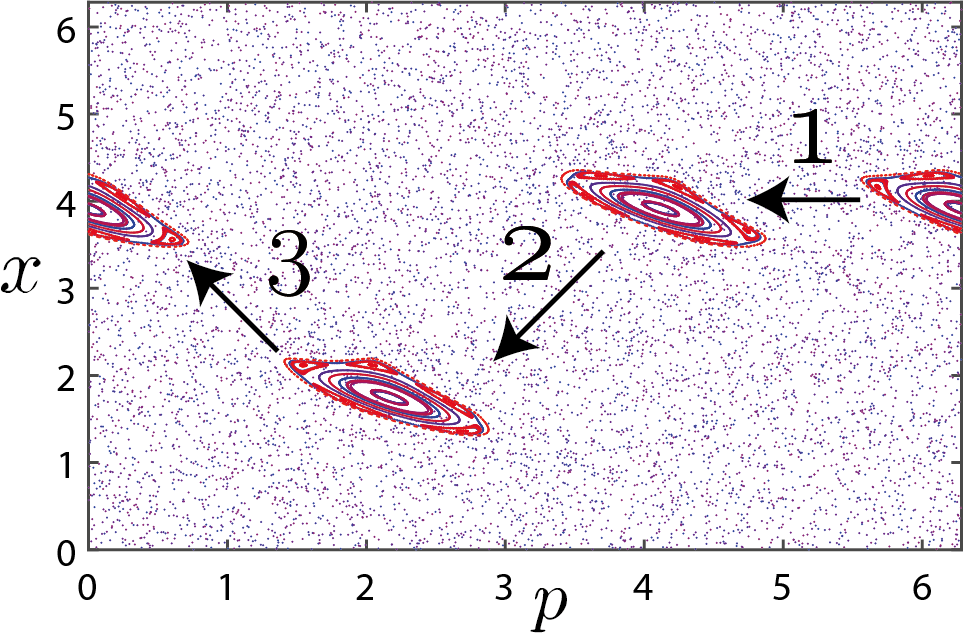}
\par\end{centering}
\medskip{}
\caption{\label{fig_comp} Left panel: Experimental, classical (blue solid
line) and quantum (red circles) momentum distributions after $15$
kicks ($\kbar=0.8$). Experimentally, the population of the peak at
the left of the plot is 14.5\%. The quantum simulation is shown as
the black dashed line. Right panel: Superposition of three successive
classical phase space structures generated by the Hamiltonian \eqref{eq_Ham}
each corresponding to 1/3 of the full period of the system. There
is thus only \emph{one} island, whose position is displayed at each
1/3 of the period. Because the classical standard map is periodic
in both $x$ and $p,$ it is possible to fold the phase space in the
$[0,2\pi[\times[0,2\pi[$ square. In the unfolded phase space, the
island is moving continuously to the left.}
\end{figure*}

 The kicked rotor is known to support so-called ``accelerator''
modes~\cite{LichtenbergLiebermann:ChaoticDynamics:82}.
However, the standard KR Hamiltonian displaying both time-reversal
and parity symmetry, if some initial condition $(x_{0},p_{0})$ leads
to classical motion in one direction, say $+x$, parity symmetry implies
that ``conjugated'' initial condition $(-x_{0},-p_{0})$ will lead
to an equivalent motion in the $-x$ direction. In this work we use
a modified Hamiltonian~\cite{Tian:EhrenfestTimeDynamicalLoc:PRB05},
that allows us to easily break parity symmetry
\begin{equation}
\hat{H}(t)=\frac{\hat{p}^{2}}{2}+K\sum_{n}\cos(\hat{x}+a_{n})\ \delta(t-n),\label{eq_Ham}
\end{equation}
where we use dimensionless units such that position and momentum operators
obey the canonical commutation relation $[\hat{x},\hat{p}]=i\kbar$,
where $\kbar$ is an effective Plank constant that can be changed
in the experiment (see below). For $a_{n}=0$, $\forall n$, we retrieve
the usual QKR Hamiltonian~\cite{Casati:LocDynFirst:LNP79}. One can
show that the following periodic phase-shift sequence $a_{n}=\varphi_{n\,{\rm [mod.}\,3]}$,
with $\{\varphi_{1},\varphi_{2},\varphi_{3}\}=\{0,2\pi/3,0\}$ produces
a breaking of the parity symmetry~\cite{Hainaut:SymmetryLocArtificialGaugeField:arXiv17}.
In the following, we will focus on relatively small values of $\kbar$,
in the range $0.8$\textendash $1.3$, allowing us to study both the
(semi-) classical dynamics and dynamical localization. A similar Hamiltonian
displaying an interplay between a quantum resonance, observed for
$\kbar=2\pi$ , and accelerator modes has been
studied experimentally~\cite{White:ExperimentalQuantumRatchet:PRA13,Sadgrove:EngineeringQuantumCorrelationsTransport:PRA13,Fishman:TheoryForQuantumAcceleratorModes:JSP18}.
In our case, in contrast, the ratchet effect is purely classical,
and is therefore independent of the value of $\kbar$.

\begin{figure*}[t!]
\centering{}\includegraphics[width=0.48\linewidth]{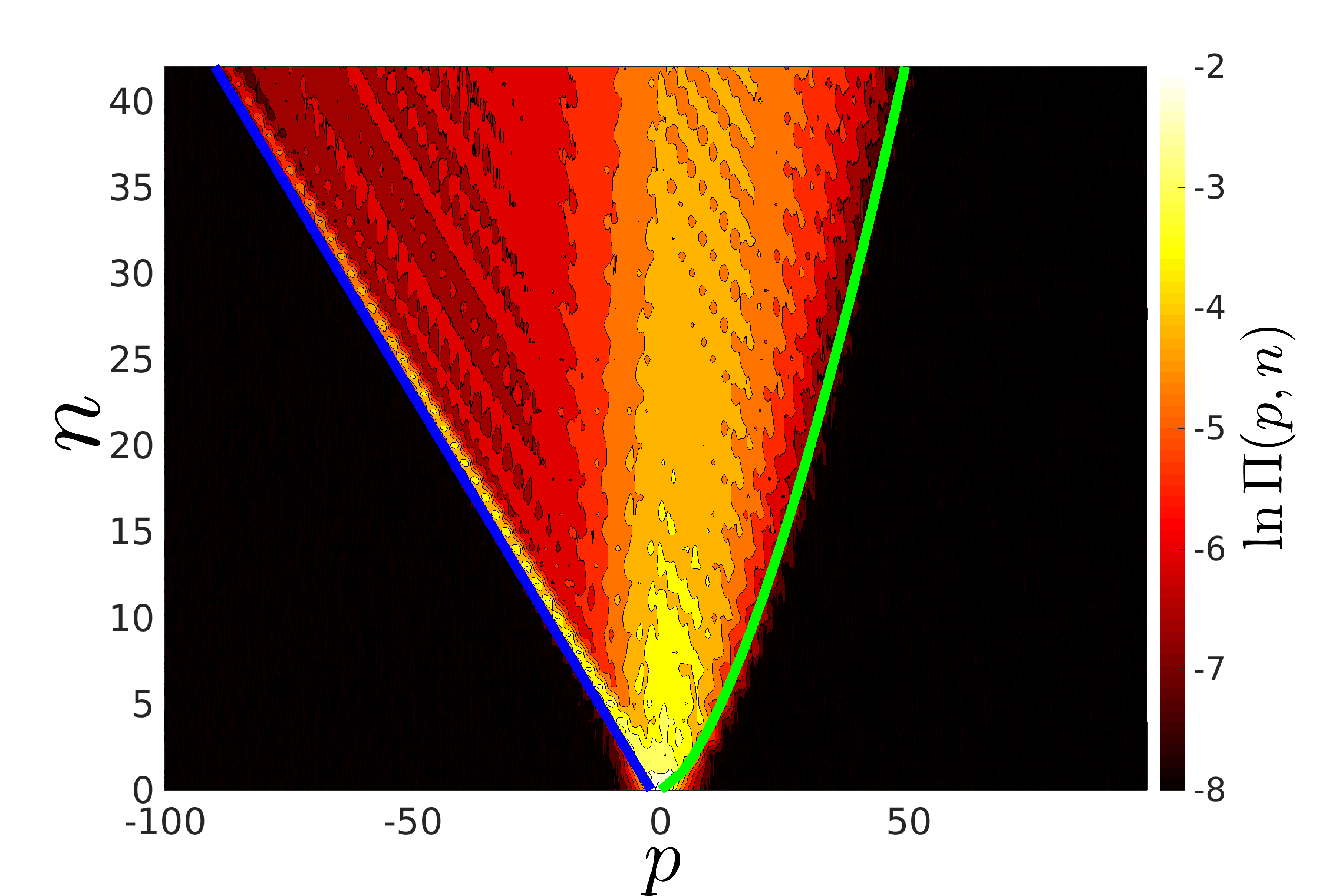}\includegraphics[width=0.46\linewidth]{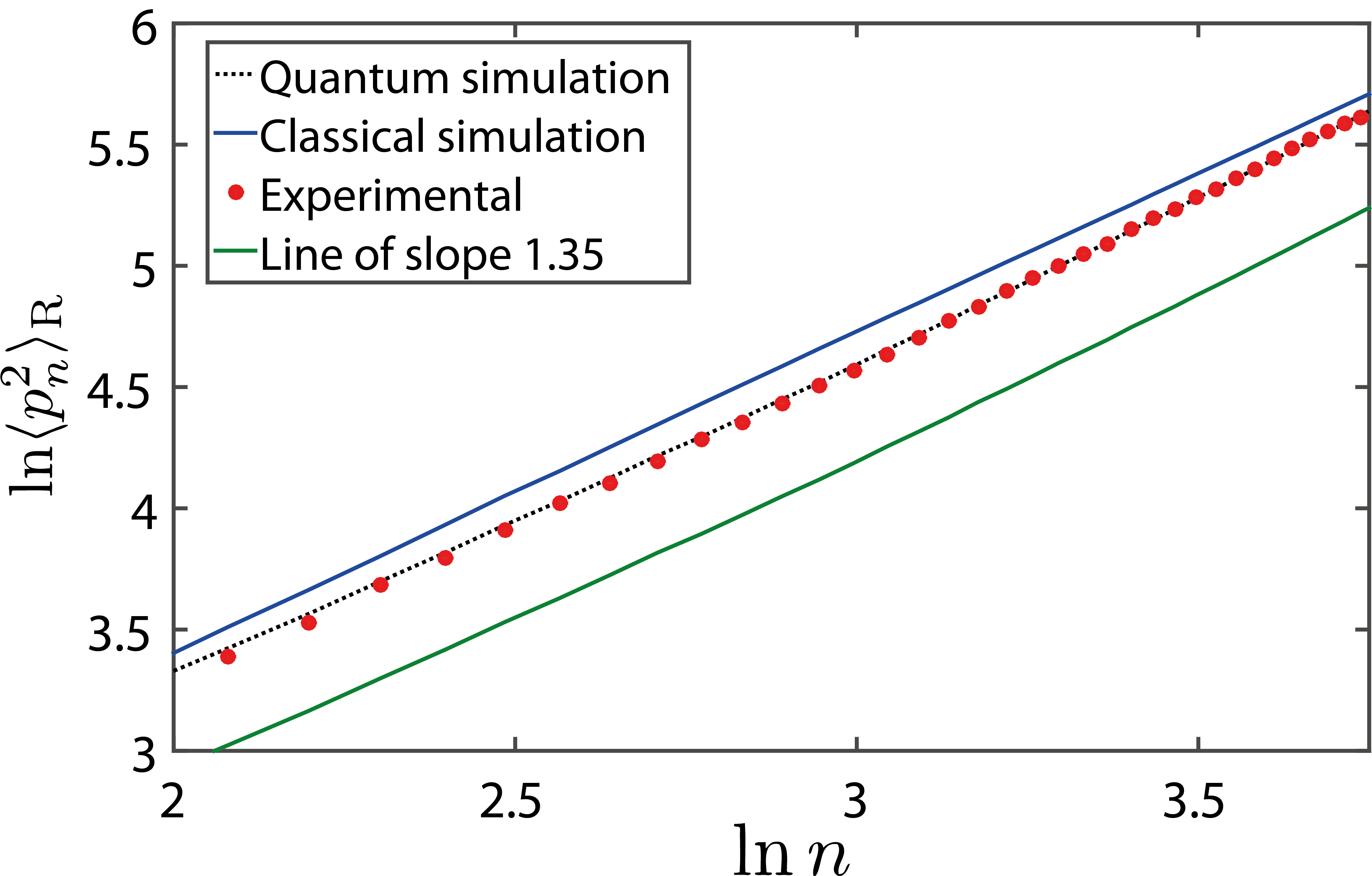}
\caption{\label{fig_cont}Left panel: False color plot of the experimental
momentum distributions as a function of momentum $p$ and of the number
of kicks $n$ ($\kbar=0.8$). The blue line on the left shows the
motion of the peak $p_{{\rm peak}}(n)=-2\pi n/3$, while the green
line on the right shows the propagation of the anomalous-diffusion
front $p_{{\rm an}}(n)\propto n^{\zeta/2}.$ Right panel: Determination
of the anomalous-diffusion exponent from the logarithmic plot of the
kinetic energy of the $p>0$ part of the system, $\langle p_{n}^{2}\rangle_{R}$,
as a function of the number of kicks $n$ ($\kbar=0.8$). The results
of the quantum and classical simulations, as well as the experimental
data, are well fitted by a power law $\langle p_{n}^{2}\rangle_{R}\propto n^{\zeta}$
with $\zeta=1.35\pm0.05$.}
\end{figure*}

Our experiment is performed with a cold ($T\simeq2.4\mu$K) cloud
of cesium atoms kicked by a periodically-pulsed, far-detuned standing
wave ($\Delta=\SI{-13}{\giga Hz}$ with respect to the \emph{D}2 line
at $\sim\SI{852}{\nano\meter}$). The beam waist is $\sim\SI{800}{\micro\meter}$
for a $\SI{330}{m\watt}$ one-beam power. The standing wave is built
by the overlap of two independent, arbitrarily-phase-modulated, laser
beams. This feature allows us to dynamically shift the potential position,
and thus to synthesize Hamiltonian \eqref{eq_Ham}. Time is measured
in units of the standing wave pulse period $T_{1}$, space in units
of $(2k_{L})^{-1}$ with $k_{L}=2\pi/\lambda_{L}$ the laser wave
vector, and momentum in units of $M/2k_{L}T_{1}$ so that $\kbar=4\hbar k_{L}^{2}T_{1}/M$
(with $M$ the atomic mass). At the maximum velocity reached by the
atoms $v_{\mathrm{max}}=\SI{0.55}{\milli/s}$, the atoms move during
the pulse duration $\tau=200$ns of a distance $v_{\mathrm{max}}\tau=110$
nm, small compared to the characteristic scale of the potential $\lambda_{L}/2=426$
nm; for most atoms one can thus consider our kicks as Dirac delta
functions. For $K\propto I/|\Delta|=3.1$ the lattice depth is about
$200$ $E_{r}$ where $E_{r}=\hbar^{2}k_{L}^{2}/2M$ is the recoil
energy. For these experimental parameters the decoherence time is
approximately 300 kicks. The main source of decoherence are the phase
fluctuations of the laser beams forming the standing wave.

The quantum dynamics of the gas can also be simulated using a Monte-Carlo
method: we choose a random initial momentum $p_{0}$ according to the distribution
\begin{equation}
|\psi_{0}(p)|^{2}=\frac{1}{(2\pi\sigma^{2})^{1/2}}e^{-\frac{p^{2}}{2\sigma^{2}}},\label{Eq:initial_density}
\end{equation}
which mimics the initial distribution of the atoms, where $\sigma=1.65\kbar$ is the width of the thermal distribution
of the atomic cloud (corresponding to $T=2.4\mu$K in the experiment).
 The plane wave $p_{0}$ is a Bloch
wave for the spatially periodic Hamiltonian (\ref{eq_Ham}), with
Bloch vector $\beta=\mathrm{frac}(p_{0}/\kbar)$, which makes it possible
to propagate it using the discrete basis set composed of eigenstates
of the momentum operator $|(m\!+\!\beta)\kbar\rangle$ with $m$ an
integer. The final momentum density is obtained by averaging the individual
momentum densities over $10^{4}$ random initial momenta. Because
$\sigma\gtrsim\kbar$ the resulting $\beta$ distribution is almost
uniform.

We first analyze the short-time dynamics for a relatively small $\kbar=0.8$
(corresponding to $T_{1}=7.67\mu$s). The left panel of Fig.~\ref{fig_comp}
compares the experimentally measured momentum distribution $\Pi(p,n)$
after $n=15$ kicks to both the classical and the quantum simulation
of the dynamics. The classical result matches very well with the experiments,
and demonstrates that the short-time dynamics is effectively classical.
The most striking feature of the momentum distribution is the sharp
peak at $p=-31.2\simeq15\times(-2\pi/3)$, which will be shown to
transport ballistically (in momentum) toward the left (negative $p$),
$p_{{\rm peak}}(n)=-2\pi n/3$. The right panel Fig.~\ref{fig_comp}
shows the classical phase portrait obtained by evolving with the classical
version of Hamiltonian \eqref{eq_Ham} $2\times10^{5}$ initial conditions
$(x_{0},p_{0})$, with $x_{0}$ uniformly sampled and $p_{0}$ sampled
according to Eq.~(\ref{Eq:initial_density}). This provides a clear
interpretation of the accelerator mode mechanism, corresponding to
the transport of a regular island across the classical phase-space.
The island's center obeys the recursion condition $(x_{n+3},p_{n+3})=(x_{n},p_{n}-2\pi)$;
as the maximum momentum transferred by one kick is $K$, changing
$|p|$ by $2\pi$ each 3 kicks requires $K\geq2\pi/3$ for the accelerator
mode to exist \cite{Ichikawa:StochasticDiffStdMap:PD87}. The value
$K\simeq3.1$ we use here allows for the largest possible island.

\begin{figure*}[t!]
\centering{}\includegraphics[width=0.5\linewidth]{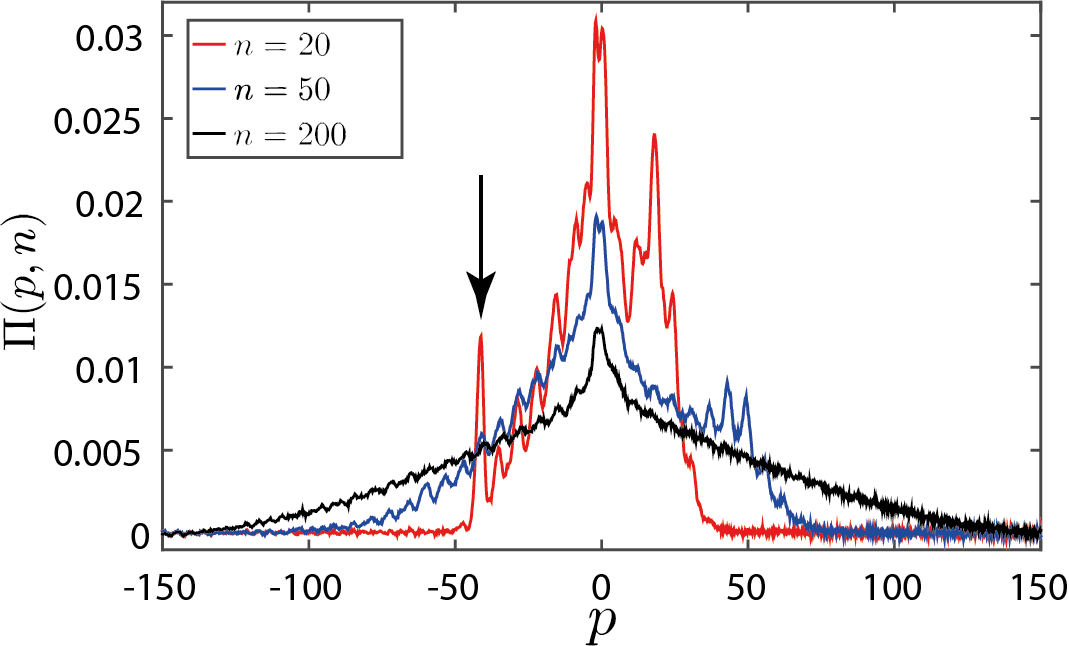}\hfill{}\includegraphics[width=0.475\linewidth]{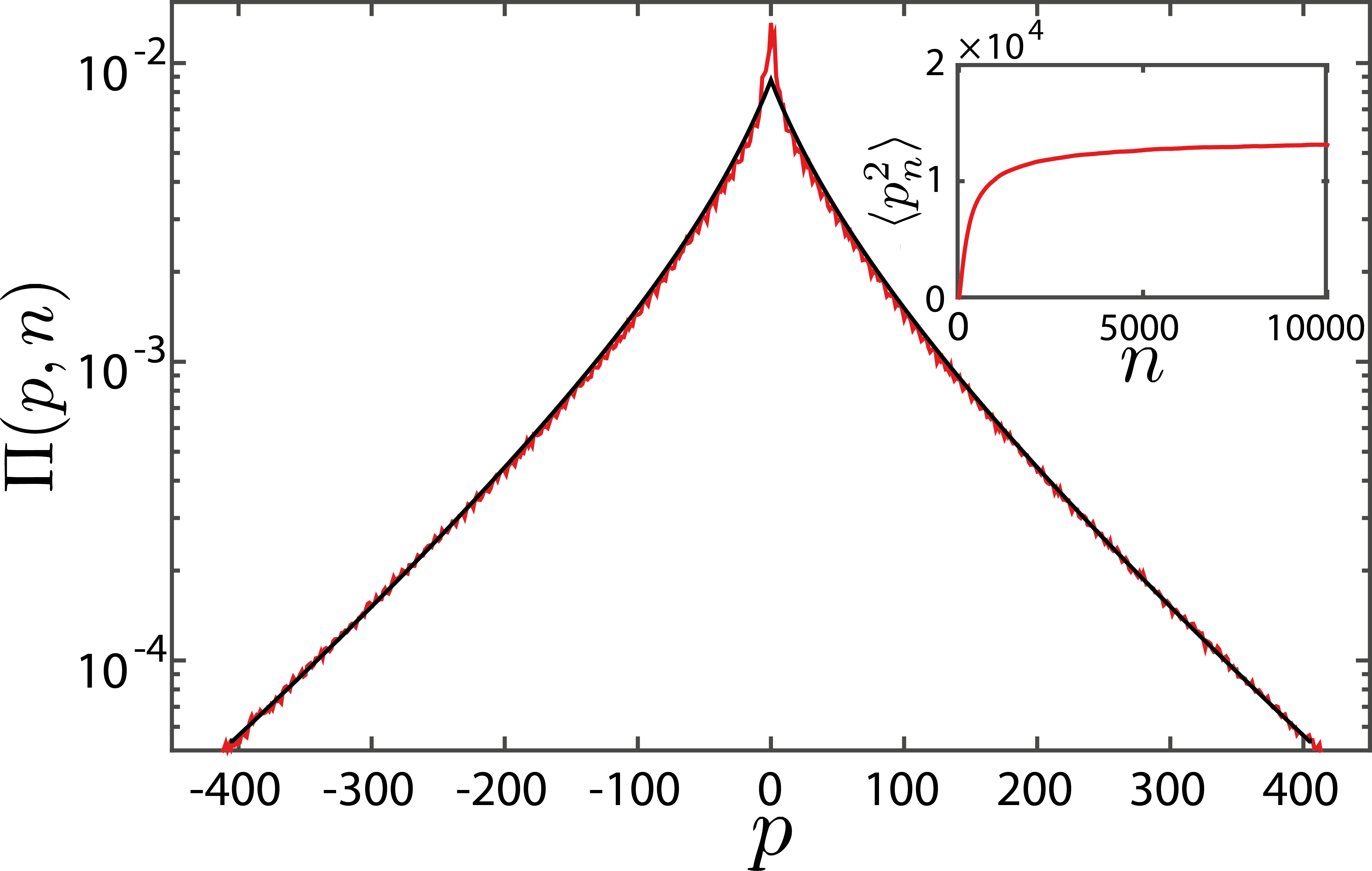}
\caption{\label{fig_resym} Left panel: Experimental momentum distributions
for $\kbar=1.3$ at three different times. The arrow shows the position
of the ballistic peak at $n=20$. At $n=200$, we observe a symmetric
momentum distribution, displaying no prominent feature. Right panel:
Simulated momentum distribution  after $n=10^{4}$ kicks (red line).
The black line is the distribution (\ref{Eq:Gogolin}), fitted with
a localization length $\xi\simeq35$. Inset : $\langle p_{n}^{2}\rangle$
as a function of the number of kicks $n$. The numerical simulation
is averaged over $5\times10^{4}$ values of the quasi-momentum $\beta$. }
\end{figure*}
Figure~\ref{fig_cont} (left) shows a false-color plot of the experimentally
measured momentum distribution $\Pi(p,n)$ as a function of both $p$
and $n$. One clearly observes the ballistic peak going to the left
(at a velocity of$-2\pi/3$ per kick), and a anomalous-diffusive front
propagating to the right. Indeed, has been shown that the presence
of accelerator modes is associated with an anomalous diffusion $\langle p_{n}^{2}\rangle\propto n^{\zeta}$,
with a non-universal exponent $\zeta\in[1,2]$~ \cite{Ishizaki:AnomalousDiffusionAccModeStdMap:PTP91,Manos:SurveyAcceleratorModesStdMap:PRE14,Gong:DirectedAnomalousDiffusionBiasedRatchet:PRE04}.
The momentum distribution is thus strongly asymmetric due to these
very different behaviors at the right and the left wings. Furthermore,
the population of the peak is seen to decrease with time. 

In contrast to the standard QKR, for which accelerator modes
always appear in counter-propagating pairs, the fact that our setup
breaks the parity symmetry allows for \textit{directed} motion. Since
unbiased classical map, such as the one studied here, cannot display
a net current when averaged over phase-space \cite{Dittrich:ClassicalAndQuantumTransportHRachet:ANPH18,Schanz:ClassicalAndQuantumHamiltonianRatchets:PRL01},
the dynamics of the chaotic sea must compensate the motion of the
ballistic peak \footnote{We have checked the vanishing of current explicitly in numerical simulations
of the classical and quantum dynamics, and it is well verified in
the experiment. The finite duration of the kicks can affect the experimental
momentum distribution at large momenta, but does not change qualitatively
our results. This has also been checked extensively by comparison
to numerical simulations.}.

Thanks to the very good resolution of our experiment, we were able
to make precise measurements of the kinetic energy of the system.
We can extract the kinetic energy contribution of positive momenta
($p>0$) $\langle p_{n}^{2}\rangle_{R}=\int_{p>0}p^{2}\Pi(p,n)\dd p$,
corresponding to the anomalously-diffusive chaotic sea, and could
thus determine by a fit the anomalous diffusion exponent $\zeta\simeq1.35\pm0.05$,
which is in good agreement with both our classical and quantum simulations,
see Fig.~\ref{fig_cont} (right). This is, to the best of our knowledge,
the first experimental evidence of the anomalous diffusion behavior
of the chaotic sea. The anomalous diffusion is usually interpreted
in terms of chaotic trajectories which spend a long time close to
the accelerated islands, thus performing L\'evy flights in the \textit{same}
direction as the island~\cite{Benkadda:SelfSimilarityStdMap:PRE97}.
Here, however, while the accelerated island propagates to the left,
the anomalous diffusion goes in the \textit{opposite} direction. Thus,
the precise mechanism responsible for this anomalous diffusion remains
an interesting open question.

For generic values of $K$ and $\kbar$, the long-time dynamics of
the QKR is governed by subtle quantum interferences which
lead to dynamical localization: asymptotically, momentum distributions
are exponentially localized. As the Hamiltonian of our system is time-periodic
(period 3) and one-dimensional, one generically expects the Floquet
states to be exponentially (Anderson) localized and the temporal dynamics
of an initially localized wavepacket to display the standard dynamical
localization at long times, possibly with a very large localization
length \cite{Shepelyansky:Kq:PD83}. In order to observe the localization
experimentally before decoherence effets become important, we have
slightly increased the value of $\kbar\simeq1.3$ (corresponding to
$T_{1}=12.46\mu$s), thus reducing the localization
time. The corresponding momentum distribution, for three different
times, are shown in, Fig.~\ref{fig_resym} (left). Note that the
population in the ballistic peak decreases
faster for this value of $\kbar$ due to the increase of quantum
tunneling out of the classical island.

We observe that the distribution gradually re-symmetrizes as it localizes.
Quite surprisingly, the ballistic side localizes first. The anomalous
diffusion front, which propagates more slowly, localizes later, but
with the same localization length. To confirm this re-symmetrization,
we have performed a  numerical simulation for the experimental parameters for times larger
than the localization time and we observed that the momentum distribution
at $10^{4}$ kicks is perfectly symmetric, see Fig.~\ref{fig_resym}
(right). The inset shows that the kinetic energy saturates due to
quantum interferences, proving dynamical localization. While this
re-symmetrization can be surprising in the context of the QKR with
directed transport, it is in fact easily understood in the context
of Anderson localization: an initially well-defined wave-packet (in
momentum) will localize and display a universal symmetric shape, called
the Gogolin distribution \cite{Efetov:SupersymmetryInDisorder:97}
\begin{equation}
\Pi_{{\rm loc}}(p)\!=\!\!\int_{0}^{\infty}\!\frac{\pi^{2}\eta\sinh(\pi\eta)\mathrm{e}^{-\frac{1+\eta^{2}}{4\xi}|p|}}{16\xi}\left(\frac{1+\eta^{2}}{1+\cosh(\pi\eta)}\right)^{2}\dd\eta.\label{Eq:Gogolin}
\end{equation}
Only the localization length $\xi$ depends on the microscopic details,
which translates here in a broad extension of the wave-packet thanks
to the ballistic mode \cite{Artuso:NonlinearPerturbationDynamicalTunneling:PRE03,Iomin:QuantumLocalizationForAKR:PRE02}.
Fig.~\ref{fig_resym} (right) also shows this universal distribution,
which indeed describes the localized regime very well. The slight
discrepancy near $p=0$ can be understood in terms of an enhanced return
to the origin effects \cite{Hainaut:EnhancedReturnOrigin:PRL17} as
well as the initial broadening of the wave-function.

In conclusion, we have shown that the careful crafting of our experimental
setup allows us to break the parity symmetry and observe a classical
Hamiltonian ratchet effect accompanied by an anomalous diffusion in
the short-time dynamics. At longer times, dynamical localization leads
to a striking re-symmetrization of the momentum distribution. It would
be interesting to better characterize the mechanism leading to an
anomalous diffusion in the direction opposite to the acceleration
mode. Also, the quantum leaking from the classical island into the
chaotic sea is important in the context of chaos assisted tunneling~\cite{Doggen:ChaosAssistedTunnelingAL:PRE17}.
This illustrates the exciting perspectives accessible when the deep
classical regime ($\kbar\ll1$) is attained experimentally. These
challenging and interesting questions are left for future work. 

The authors thank G. Lemari\'e and N. Cherroret for fruitful discussions.
This work is supported by Agence Nationale de la Recherche (Grant
K-BEC No. ANR-13-BS04-0001-01), the Labex CEMPI (Grant No. ANR-11-LABX-0007-01),
and the Ministry of Higher Education and Research, Hauts de France
Council and European Regional Development Fund (ERDF) through the
Contrat de Projets Etat-Region (CPER Photonics for Society, P4S).

%
\end{document}